\documentclass{aa}  

\usepackage[switch]{lineno}
\usepackage{graphicx}
\usepackage{natbib}
\usepackage{txfonts}
\usepackage[draft]{hyperref}

\usepackage[usenames]{color}

\begin{document} 

% add line numbers to plot
%\linenumbers
%\modulolinenumbers[5]

   \title{Adaptive elliptical aperture photometry: a software package for high-cadence ground-based photometry\thanks{The {\sc TEA-Phot} software can be downloaded from: \url{https://bitbucket.org/DominicBowman/tea-phot/}}}
   \titlerunning{the {\sc TEA-Phot} pipeline}
   
   \subtitle{I. Application to rapid oscillators observed from SAAO\thanks{Based on observations made using Sutherland High Speed Optical Camera (SHOC) on the 1-m telescope at South African Astronomical Observatory (SAAO), Sutherland, South Africa.}}
   %\subtitle{}

   \author{Dominic M. Bowman \inst{1} % \fnmsep\thanks{footnote 1}
          \and
          Daniel L. Holdsworth \inst{2,3}
          }

    \institute{Institute of Astronomy, KU Leuven, Celestijnenlaan 200D, 3001 Leuven, Belgium \\
              \email{dominic.bowman@kuleuven.be} %\thanks{footnote 2}
         \and
         	Jeremiah Horrocks Institute, University of Central Lancashire, Preston PR1 2HE, UK 
	\and
		Center for Space Research, North-West University, Mafikeng Campus, Private Bag X2046, Mmabatho 2745, South Africa
	 }

   \date{Received 8 April 2019 / accepted 5 July 2019}

% \abstract{}{}{}{}{} 
% 5 {} token are mandatory
 
   \abstract
   {Modern space telescopes are currently providing high-precision light curves for a large fraction of the sky, such that many new variable stars are being discovered. However, some stars have periodic variability with periods of order minutes and require high-cadence photometry to probe the physical mechanisms responsible. A cadence of less than a minute is often required to remove Nyquist ambiguities and confirm rapid variability which forces observers to obtain high-cadence ground-based photometry.}
   {We aim to provide a modern software package to reduce ground-based photometric time series data and deliver optimised (differential) light curves. To produce high-quality light curves, which maximise the amplitude signal-to-noise ratio of short-period variability in a Fourier spectrum, we require adaptive elliptical aperture photometry as this represents a significant advantage compared to aperture photometry using circular apertures of fixed radii.}
   {The methodology of our code and its advantages are demonstrated using high-cadence ground-based photometry from the South African Astronomical Observatory (SAAO) of a confirmed rapidly oscillating Ap (roAp) star. Furthermore, we employ our software package to search for rapid oscillations in three candidate roAp stars.}
   {We demonstrate that our pipeline represents a significant improvement in the quality of light curves, and we make it available to the community for use with different instruments and observatories. We search for and demonstrate the lack of high-frequency roAp pulsations to a limit of ~$\sim1$~mmag using $B$ data in the three Ap stars HD~158596, HD~166542 and HD~181810.}
   {We demonstrate the significant improvement in the extraction of short-period variability caused by high-frequency pulsation modes, and discuss the implication of null detections in three Ap stars.}

   \keywords{techniques: photometric-- stars: chemically peculiar -- stars: oscillations -- stars: early-type -- stars: individual: HD~158596; HD~166542; HD~181810; J1640 }

   \maketitle

%%%%%%%%%%%%%%%%%%%%%%%%%%%%%%%%%%%%%%%%%%%%%%%%%%

\section{Introduction}
\label{section: introduction}

In the study of variable stars using time series photometry, it is always the goal to strive for high-precision in both time and stellar flux. Modern telescopes and CCDs have allowed a vast array of stellar variability to be probed, and this is especially true for space telescopes such as CoRoT \citep{Auvergne2009}, {\it Kepler} \citep{Borucki2010}, and K2 \citep{Howell2014}. These missions have provided high-quality light curves with unprecedented photometric precision for hundreds of thousands of stars. However, despite the drastic increase in the number of stars with space photometry, short-period phenomena, such as rapidly oscillating stars with periods of order minutes, require a high cadence of order tens of seconds to accurately probe the variability. 

Specific astrophysical examples for which high-cadence photometry is typically required include the rapidly oscillating Ap stars (roAp; \citealt{Kurtz1978c, Kurtz1982c, Holdsworth2018d}), pulsating white dwarfs (e.g. \citealt{Fontaine2008e, Winget2008, Hermes2017f}), and pulsating sub-dwarfs (e.g. \citealt{Charpinet2006c, Ostensen2010a, Holdsworth2017a}). We refer the reader to \citet{ASTERO_BOOK} for a review of pulsational variability across Hertzsprung--Russell (HR) diagram and the associated time scales. Inevitably, a high cadence of order 1~min is expensive for space telescopes in terms of onboard data storage and/or downlink speeds. Therefore, this has resulted in a dearth of high-cadence photometric data available for short-period variable stars, which is particularly pertinent for the roAp stars.

To secure the necessary time series photometry of pulsating stars with a cadence of order tens of seconds, astronomers typically obtain ground-based photometry to search for, confirm and characterise short-period variability in stars caused by pulsations. However, obtaining high-quality ground-based time-series photometry remains a particularly challenging task when studying variable stars because of the various seeing timescales present in the Earth's atmosphere. In variable star studies it is typically the amplitude signal-to-noise ratio (S/N) of a peak in a Fourier transform that should be maximised, since the pulsation frequencies and corresponding amplitudes represent the fundamental data for any subsequent analysis \citep{ASTERO_BOOK}. However, even under photometric conditions, the noise level in a Fourier spectrum and the S/N of pulsation mode frequencies depend on the data reduction and photometry pipeline. We refer the reader to \citet{Howell2006a} for a detailed discussion of CCDs and \citet{Kurtz2000c} for the problems that can occur that affect any resultant S/N-based criterion for detecting pulsations in a light curve.

A prime example for which high-cadence ground-based photometry is often needed is for the roAp stars --- see \citet{Kurtz1982c} for a review of roAp stars. In the region of the HR~diagram where roAp stars are located, i.e. where the main sequence and classical instability strip intersect, there is a complex relationship amongst the physics of rotation, pulsation, binarity and magnetic fields. Approximately 10\,\% of these stars are classified as Ap (or Bp; CP\,2) stars \citep{Wolff1968a, Preston1974, Power2007, Sikora2018a}, which have long rotation periods of order days, but can be as short as 0.5~d (e.g. see \citealt{Adelman2002d, Mathys2004a, Mathys2015}) or longer than a century (e.g. \citealt{Mathys2015}). The Ap stars host a strong large-scale magnetic field of order 1~kG (e.g. \citealt{Babcock1960, Auriere2004, Buysschaert2018b}), which is thought to be fossil in origin and be efficient to brake Ap stars on the pre-main sequence \citep{Abt1995, Stepien2000}. The slow rotation and strong magnetic field allow atomic diffusion and gravitational settling to separate chemical elements in the outer layers of the stellar envelope, and produce surface abundance inhomogeneities which modulate the light curve as the star rotates with respect to an observer \citep{Stibbs1950}.

The roAp stars pulsate in high-overtone pressure (p) modes with periods that range from 4.6 to 24~min \citep{Kurtz1982c, Kurtz1990a, Martinez1994f, Alentiev2012, Smalley2015, Joshi2016, Cunha2019a}, and have photometric peak-to-peak light curve amplitudes as large as 34~mmag in Johnson $B$ data \citep{Kurtz1990a, Holdsworth2018a}. Fewer than 80 roAp stars are known, so more are needed to accurately probe the observational instability region of these rare stars and understand the interior properties by means of their pulsations (see e.g., \citealt{Cunha2002d, Cunha2013, Cunha2019a, Holdsworth2018b}). Recently, \citet{Bowman2018b} performed a systematic search for pulsations in 56 ApBp stars observed by the K2 space mission \citep{Howell2014}, and identified three of these stars, HD~158596, HD~166542 and HD~181810, as candidate roAp stars. Since the 29.5-min cadence of the K2 mission defines a Nyquist frequency of 24.47~d$^{-1}$ and because low-overtone p-mode pulsations are not predicted to be excited in chemically peculiar stars with magnetic field strengths of order 1\,kG \citep{Saio2005, Saio2014a}, it was suggested that these isolated peaks may be Nyquist alias frequencies of high-frequency roAp pulsation modes \citep{Bowman2018b}.

To definitively confirm an Ap star as a roAp, high-cadence photometry is needed, which is normally obtained using ground-based telescopes. To analyse such a data set, there have been a selection of image reduction and aperture photometry software packages made available to the astronomy community in recent decades, including {\sc DoPHOT} \citep{Schechter1993} and {\sc daophot} \citep{Stetson1987c} and Image Reduction and Analysis Facility (IRAF) {\sc apphot} package \citep{Tody1986, Tody1993}. In this paper, we describe our \texttt{PYTHON} adaptive elliptical aperture photometry ({\sc TEA-Phot}) pipeline and demonstrate its flexibility and advantages for South African Astronomical Observatory (SAAO) photometry over other commonly used routines using fixed circular apertures in section~\ref{section: method}. Specifically, we demonstrate the marked advantages of this approach using a known roAp star studied by \citet{Holdsworth2018b} using high-cadence ground-based photometric observations from the SAAO. In section~\ref{section: results}, we apply our pipeline to SAAO photometry of the three candidate roAp stars HD~158596, HD~166542 and HD~181810 discussed by \citet{Bowman2018b} to investigate if they exhibit high-frequency pulsation modes. Finally we conclude in section~\ref{section: discussion}.

% % % % % % % % % % % % % % % % % % % % % % % % % 

\section{The {\sc TEA-Phot} pipeline: adaptive elliptical aperture photometry}
\label{section: method}

The current pipeline to reduce and analyse SAAO photometric time series available to observers is based on the {\sc DoPHOT} software \citep{Schechter1993}, and employs \texttt{IRAF} \citep{Tody1986, Tody1993} to perform image reduction and aperture photometry using circular apertures of fixed radii. Although \texttt{IRAF} was a pioneering software package and has been very successful within astronomy, its flexibility, suitability and accuracy in light of today's high-precision observations is limited. For example, stellar magnitudes are not output to a high-enough precision for studying stars with pulsation mode amplitudes of order 1~mmag. Later updated options for image reduction, including for example \texttt{PyRAF} are available, but even this is no longer supported on the latest computer hardware at Space Telescope Science Institute (STScI; \citealt{Lucas2018}).

In this work, it is our goal to provide a high-quality differential photometry pipeline to the community, which in its current form is optimised for high-cadence ground-based observations of roAp stars observed at SAAO. However, it can be easily re-purposed for other scientific goals and/or modified for use with other instruments and observatories. To maximise our prospects, we have developed a customised pipeline in \texttt{PYTHON} for the reduction and processing of SAAO photometry utilising the functionality of the \texttt{SEP}\footnote{SEP: \url{https://sep.readthedocs.io/en/v1.0.x/}} module \citep{Barbary2016}. The \texttt{SEP} module makes the algorithms of the Source Extractor library \citep{Bertin1996} available as stand-alone functions within a \texttt{PYTHON} environment, thus offering a great range of flexible options for performing aperture photometry. An important advantage when using SEP, whose functionality is based on the original Source Extractor software, is its ability to perform {\it elliptical} aperture photometry --- see e.g., \citet{Mallonn2018a} and references therein. For high-cadence ground-based photometry of pulsating stars, this is significant improvement over the standard approach to use circular apertures of fixed radii, as it takes into account the possible asymmetric affects of seeing in the atmosphere. Furthermore, \texttt{SEP} allows {\it adaptive} aperture photometry using ellipses, such that the optimum aperture's dimensions can be determined for each individual CCD frame.

The flexibility of the \texttt{PYTHON} \texttt{SEP} module allowed us to incorporate it into a standalone \texttt{PYTHON} pipeline for reducing and extracting (differential) light curves from SAAO photometry. In subsection~\ref{subsection: pipeline} we briefly outline the functionality and flexibility of our adaptive elliptical aperture photometry ({\sc TEA-Phot}) pipeline, and demonstrate its advantages by comparing it to the {\sc DoPHOT} pipeline \citep{Schechter1993} using the previously-known roAp star J1640 studied by \citet{Holdsworth2018b} in subsection~\ref{subsection: J1640}.

%	%	%	%	%	%	%	%	%	%	%	%	%	

	\subsection{{\sc TEA-Phot} pipeline overview}
	\label{subsection: pipeline}	
	
	In its released format, our {\sc TEA-Phot} pipeline is optimised for photometry using either of the two available Sutherland High Speed Optical Camera (SHOC; \citealt{Coppejans2013}) instruments (SHA or SHD)\footnote{SHOC documentation: \url{http://www.saao.ac.za/science/facilities/instruments/shoc/}}, which output single \texttt{FITS} cubes for each observing block. However, other instruments that output an individual \texttt{FITS} file per exposure can also be used, with functionality for the STE3 CCD camera available at SAAO already included as an option when using {\sc TEA-Phot}. For the following discussion, we describe the use of {\sc TEA-Phot} with SHOC photometry, but the methodology is the same if used with other instruments.
	
	For basic use, one simply has to specify the name of the \texttt{FITS} cube of the images from the command line:
	
	\vspace{0.5cm}
	\noindent \texttt{> python TEA-Phot.py <obs> <inst> <{image$\_$cube}>}
	\vspace{0.5cm}
	
	\noindent where \texttt{<obs>}, \texttt{<inst>} and \texttt{<image$\_$cube>} are the names of the observatory and instrument, and the filename of the image \texttt{FITS} cube to be analysed, respectively. The {\sc TEA-Phot} code can be used with different instruments at different observatory locations, but it is currently optimised for use with the SHOC instrument at SAAO. Other optional command-line arguments are available, such as flags to perform image reduction (bias subtraction and flat fielding), sigma clipping of outliers and atmospheric extinction correction to the resultant individual star light curves. The full list of optional command line arguments and their descriptions can be found using:
	
	\vspace{0.5cm}
	\noindent \texttt{> python TEA-Phot.py \textendash\textendash help}
	\vspace{0.5cm}
		
	The {\sc TEA-Phot} pipeline is user friendly and interactive as it allows the user to simply click on the \texttt{MATPLOTLIB} image to select the star(s) of interest. When used to also perform image reduction, individual bias and flat-field frames are extracted and combined into master bias and flat frames, respectively. Each image frame is corrected using using the standard procedure of subtracting the master bias frame and normalising by the master flat field frame. Image reduction is not necessary to run {\sc TEA-Phot}, but is strongly advised for creating light curves.
	
\begin{figure}[t] 
	\centerline{\includegraphics[width=0.99\columnwidth]{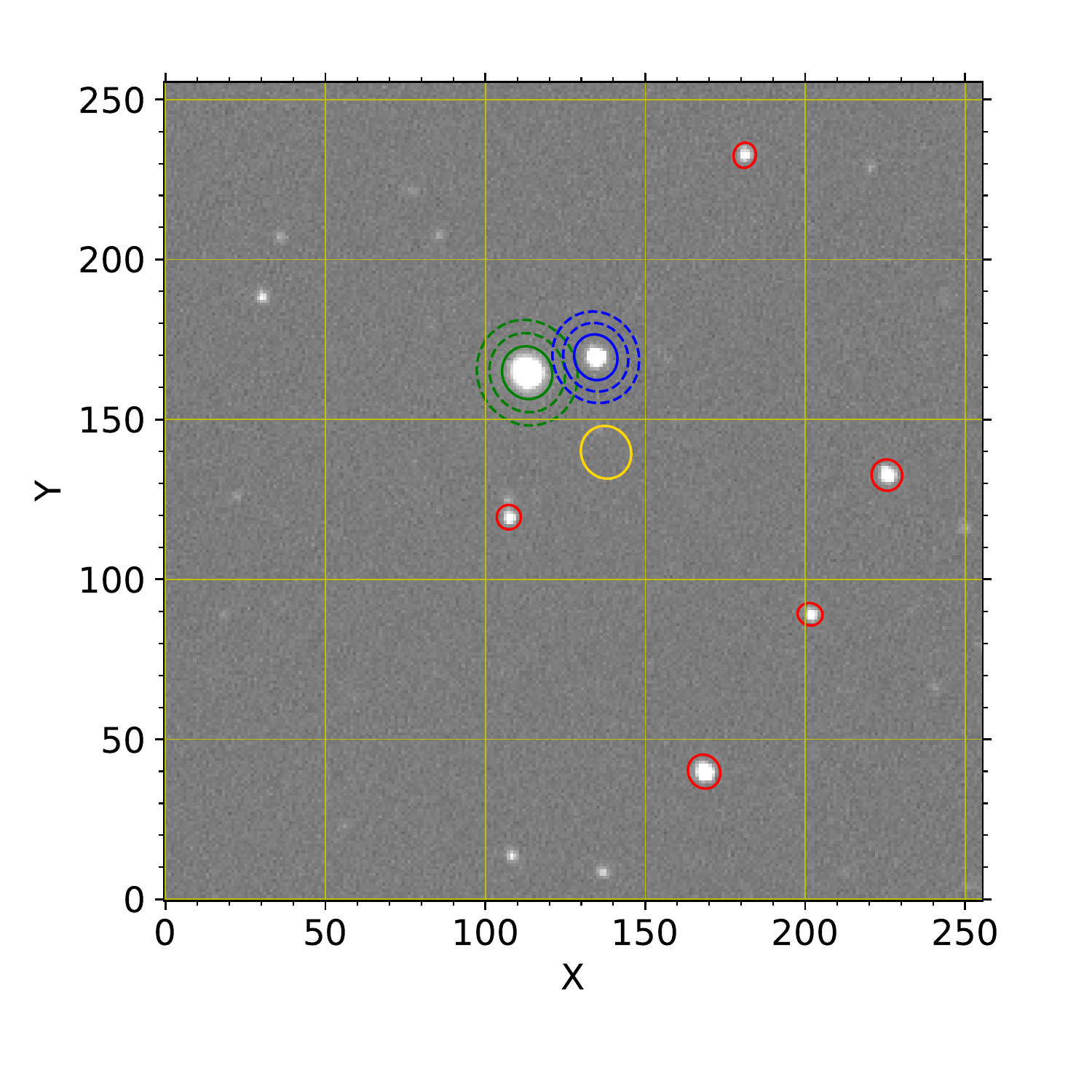}}
	\caption{Example of a reduced CCD image frame with extracted sources highlighted with red ellipses, and the target and comparison stars in green and blue, respectively. Note that the ellipses shown here do not represent the optimum apertures, but are only for illustrative purposes. The dashed lines around the target and comparison stars represent the annuli that can be used to calculate the sky background, whereas the yellow ellipse indicates the location of a {\it `sky star'} that can also be used to calculate and subtract the sky background.} 
	\label{figure: labelled sources}
	\end{figure}

	Source extraction is performed for each image using \texttt{SEP} with a default S/N criterion of 10, but can be defined by the user as a command-line argument. The interactive nature of the {\sc TEA-Phot} pipeline allows the user to select the target and comparison stars using the first image frame in an observed \texttt{FITS} cube. In Fig.~\ref{figure: labelled sources}, an example of a reduced image in a given cube is provided, in which all extracted sources using S/N~$\geq 10$ are indicated using red circles, and the user-selected target and comparison stars shown in green and blue circles, respectively.
		
\begin{figure}[t] 
	\centerline{\includegraphics[width=0.99\columnwidth]{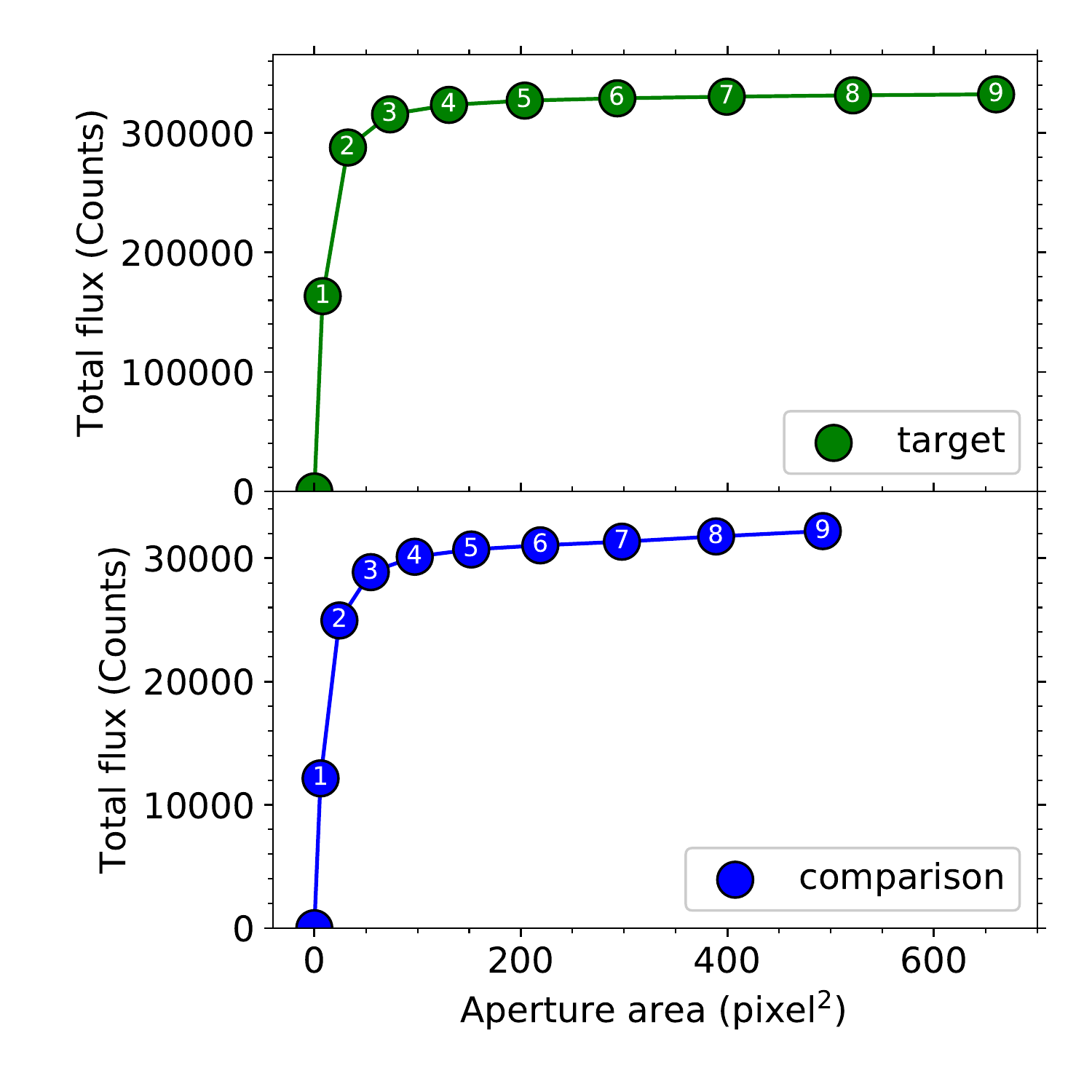}}
	\caption{Curve of growth for total flux versus aperture size for the target star in green and the comparison star in blue.}
	\label{figure: curve of growth}
	\end{figure}

	Inspired by \citet{Overcast2010}, we use the {\it curve of growth} method within {\sc TEA-Phot} to determine the optimum major- and minor-axis dimensions of the elliptical apertures to be used to extract light curves. The user is able to directly visualise the appropriate aperture size using the curve of growth for the target and comparison stars, as demonstrated in Fig.~\ref{figure: curve of growth}. For example, selecting an aperture larger than the {\it `fifth'} option does not significantly increase the target flux, yet does increase the sky background contribution. Of course, the choice of optimum aperture depends on the observing conditions, such as seeing, hence the need for flexibility in aperture size from frame-to-frame and from night-to-night. The user's choice of optimum aperture size is then used as an initial input within \texttt{SEP} for determination of the optimum aperture in each frame and the subsequent extraction of target and comparison star light curves.
	
	For the determination and subtraction of the sky background, two options are available for the user to choose from. The first option is to use an annulus around each star's location, which is defined as having minimum and maximum dimensions of 1.5 and 2 times larger than a star's optimum aperture, respectively. The second option available is to choose the location of a {\it `sky star'}, such that the sky background is calculated using the optimum aperture of each star but at a location on the CCD image frame in which there are no sources as chosen by the user. The latter of these two options is especially useful for crowded fields in which the annuli option may include other nearby sources. An example of the two options is shown in Fig.~\ref{figure: labelled sources}, in which the dashed-line ellipses around the target and comparison stars show the inner and outer edges of the annuli sky background option, whereas the yellow ellipse shows the user-defined location of a {\it `sky-star'} to be used for the sky background measure.
	
	Aperture photometry is performed for all image frames with the final time series results output as an \texttt{ascii} file. The output includes the midpoint of each exposure, which are the Coordinated Universal Time (UTC) time stamps provided in the \texttt{FITS} header converted into Heliocentric Julian Date (HJD$-2450000.0$), and also to Barycentric Dynamical Time (TDB) expressed in Barycentric Julian Date (BJD$-2450000.0$). The conversion into HJD (UTC) and BJD (TDB) is done using the \texttt{PYTHON} \texttt{ASTROPY} libraries \citep{Astropy2013} using the date, time, stellar co-ordinates and the observatory's latitude, longitude and altitude, which are also provided in the \texttt{FITS} header. The differential (target $-$ comparison), and individual target and comparison star light curves are also provided. It is possible to include an atmospheric extinction and airmass correction to the individual target and comparison stars' light curves if desired by the user, but this requires the extinction coefficient for the filter and observatory site to be known {\it a priori} and included as a command-line input.

%	%	%	%	%	%	%	%	%	%	%	%	%	

	\subsection{Application to the roAp star J1640}
	\label{subsection: J1640}
	
	Here we use the published case study of the roAp star J1640 (2MASS~16400299-0737293) to demonstrate the significant advantages of our {\sc TEA-Phot} pipeline over the current {\sc DoPHOT} pipeline available to the community to reduce SAAO photometry. The roAp star J1640 has a spectral type of A7\,Vp\,SrEu(Cr) and its high-amplitude dominant pulsation mode has a frequency of $\nu = 151.93$~d$^{-1}$ (i.e. period of 9.5~min). This star was previously studied by \citet{Holdsworth2018b}, to which we refer the reader for a more detailed analysis of the complete SAAO light curve.
	
\begin{figure*}[t] 
	\centerline{\includegraphics[width=0.999\textwidth]{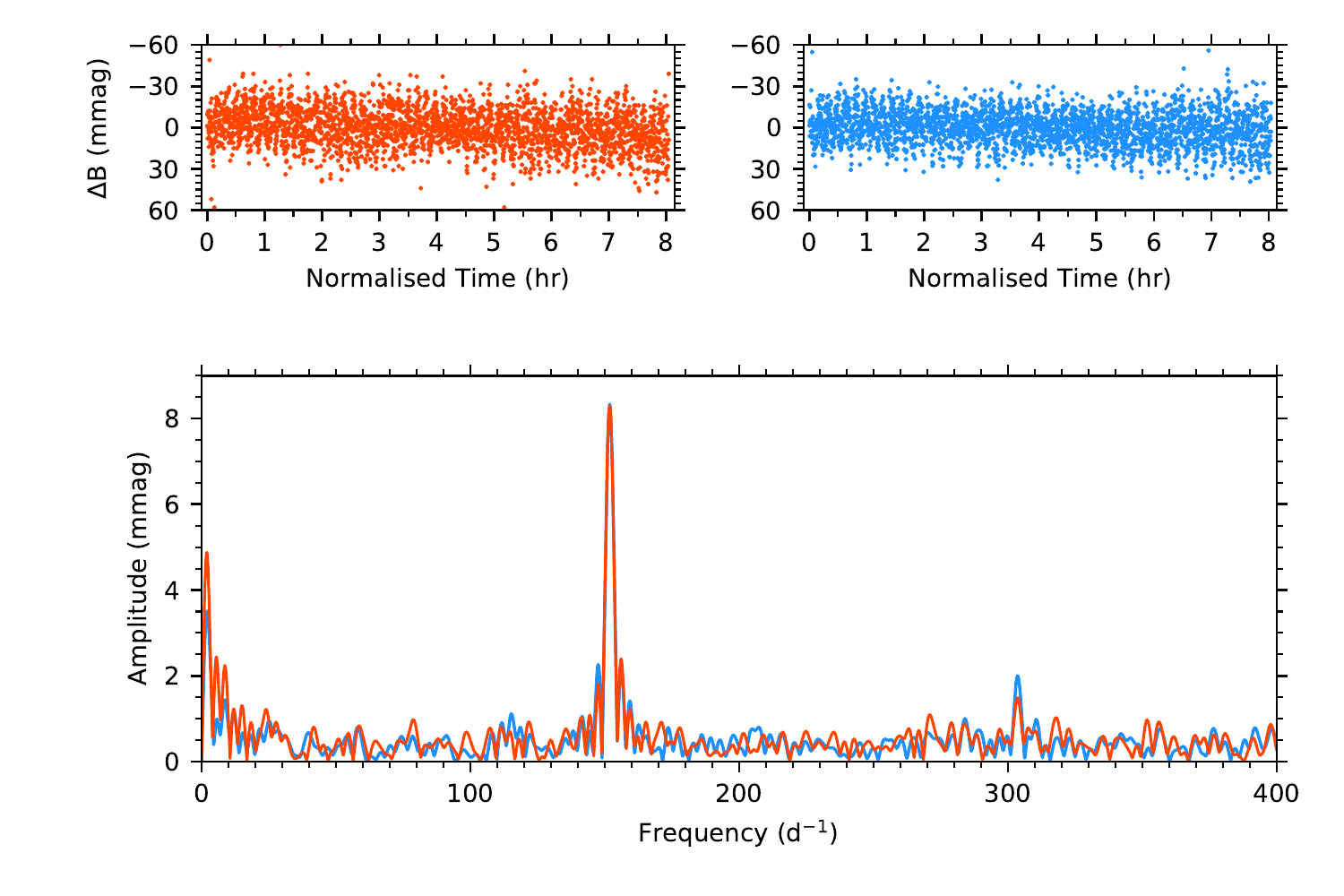}}
	\caption{The differential light curves of the roAp star J1640 extracted using the {\sc DoPHOT} pipeline (red; top-left) and with the our {\sc TEA-Phot} pipeline (blue; top-right). These data were obtained using the SHOC instrument at SAAO on 12 June 2017. The bottom panel demonstrates the difference in the S/N of the pulsation mode at $151.93$~d$^{-1}$ and its harmonic at $303.86$~d$^{-1}$ in the amplitude spectra of these two light curves.}
	\label{figure: J1640}
	\end{figure*}

	In our comparison, we compare the light curves extracted by the {\sc DoPHOT} and our {\sc TEA-Phot} pipelines using 8-hr of $B$ data obtained on 12 June 2017, which consists of 2859 individual frames with an exposure time of 10~sec that were obtained using the SHA SHOC instrument mounted on the SAAO 1.0-m telescope. The extracted {\sc DoPHOT} and {\sc TEA-Phot} light curves are shown in red and blue in the top-left and top-right panels of Fig.~\ref{figure: J1640}, respectively. We calculated Discrete Fourier transforms (DFT; \citealt{Deeming1975, Kurtz1985b}) and show the amplitude spectra of the light curves obtained from the {\sc DoPHOT} and {\sc TEA-Phot} pipelines in the bottom panel of Fig.~\ref{figure: J1640}, in which the amplitude and therefore S/N of the dominant pulsation mode frequency (and its harmonic at 303.86~d$^{-1}$) is significantly higher when using our {\sc TEA-Phot} pipeline. The noise level at high-frequency is also lower in the amplitude spectrum obtained using {\sc TEA-Phot} because of the smaller variance in the light curve. We provide the S/N values of the pulsation mode and its harmonic in both data sets in Table~\ref{table: SNR} as well as the root mean square (RMS) of the extracted light curves, demonstrating the improved values obtained using the {\sc TEA-Phot} pipeline compared to {\sc DoPHOT}. 
	
	We obtain similarly improved results for all available SAAO photometry of the roAp star J1640 when using {\sc TEA-Phot}, especially when observing in non-photometric conditions. The improvement in the ability to extract high-frequency pulsation modes in the light curves of variable stars extracted using our pipeline is clear, with adaptive elliptical apertures better suited to compensate for non-negligible atmospheric seeing that varies on time scales similar to the short integration times of order 10~sec needed to study high-frequency pulsators.

	% - - - - - % - - - - - % - - - - - % - - - - - %
	\begin{table}
	\caption{The S/N values of the dominant pulsation mode (151.93~d$^{-1}$) and its harmonic (303.86~d$^{-1}$) using the amplitude spectra, and the root mean square (RMS) of the differential photometry with the {\sc DoPHOT} and {\sc TEA-Phot} pipelines for the roAp star J1640.} 
	\begin{center}
	\begin{tabular}{l c c c}
	\hline \hline
	\multicolumn{1}{c}{Pipeline} & \multicolumn{2}{c}{Amplitude S/N} & \multicolumn{1}{c}{RMS} \\
	{} & {pulsation} & {harmonic}  & {(mmag)} \\ \hline
	{\sc DoPHOT}	&	7.09		&	2.54		&	20.85	\\
	{\sc TEA-Phot}	&	7.38		&	3.81		&	12.77	\\
	\hline \hline
	\end{tabular}
	\end{center}
	\label{table: SNR}
	\end{table}
	% - - - - - % - - - - - % - - - - - % - - - - - % 
	
%	%	%	%	%	%	%	%	%	%	%	%	%	

	\subsection{Source code}
	\label{subsection: source code}
	
	We are committed to open science, and have made our \texttt{PYTHON} {\sc TEA-Phot} pipeline publicly available via \url{https://bitbucket.org/DominicBowman/TEA-Phot/}. The code is fully documented and is provided under a GPL~v3 license. Its dependencies are \texttt{ASTROPY} \citep{Astropy2013}, \texttt{MATPLOTLIB} \citep{Hunter2007}, and \texttt{SEP} \citep{Barbary2016}, all of which are freely available and installable using the \texttt{conda} and \texttt{pip} package managers.
	
	 We encourage those who are interested for scientific (and outreach) purposes to use our code for extracting light curves obtained by a SHOC instrument at SAAO. Furthermore, our code can be used a starting point for extracting light curves from other instruments, but a word of caution is that each instrument, telescope and observatory have different keywords and file formats, therefore it is necessary to modify the code where appropriate to overcome this. In a future release, the {\sc TEA-Phot} code will be expanded to include other instruments, such as the MAIA three-channel imager mounted on the 1.2-m Flemish Mercator telescope on La Palma \citep{Raskin2013}.

% % % % % % % % % % % % % % % % % % % % % % % % % 

\section{Results: ground-based photometry of three candidate roAp stars discovered by K2}
\label{section: results}

In this section, we investigate the hypothesis made by \citet{Bowman2018b} that the three ApBp stars, HD~158596, HD~166542 and HD~181810, are candidate roAp stars. The known properties of these stars are given in Table~\ref{table: stars}, which includes their $B$ and $V$ magnitudes from \texttt{SIMBAD}, spectral type from \citet{Renson2009}, rotation period from \citet{Bowman2018b} and polar magnetic field strength (if available) from \citet{Buysschaert2018b}. The claim that each of HD~158596, HD~166542 and HD~181810 was a candidate roAp star was made based on the detection of a significant isolated peak in the amplitude spectra of K2 data of these stars, which was interpreted to be a possible Nyquist alias frequency of a high-frequency roAp pulsation mode. However, the Nyquist frequency of these K2 data was only 24.47~d$^{-1}$, which was insufficient to confirm the origin of these frequencies in these stars.

To confirm if HD~158596, HD~166542 and HD~181810 are roAp stars, we obtained high-cadence ground-based $B$ data using the SHOC instrument \citep{Coppejans2013} mounted on the 1.0-m telescope at the Sutherland site of the SAAO in May and June of 2018. A Johnson $B$ filter was used for all observations as the S/N ratio of roAp pulsations is largest within this wavelength range \citep{Medupe1998b}. A summary of the observations used in our analysis is given in Table~\ref{table: observations}. Each star was observed for a minimum of 2~hr with an exposure time of either 5 or 10~sec, such that the resultant Nyquist frequency is much higher than the frequency range of roAp pulsation modes --- i.e. $60 \lesssim \nu \lesssim 300$~d$^{-1}$.

We applied our {\sc TEA-Phot} pipeline, binned all the data of the same star to have the same exposure time and calculated amplitude spectra of the extracted light curve to search for high-frequency roAp pulsation modes. Since ground-based photometry obtained over more than a single night suffers from significant aliasing effects and since a time series from only a single night has a poor frequency resolution, we are unable to probe low-frequency variability (defined here as $\nu \lesssim 50$~d$^{-1}$) in the data of our stars. Owing to the oblique pulsator model associated with roAp stars \citep{Kurtz1982c}, amplitude modulation of pulsations occurs during the rotation period. This is because the pulsation axis is typically inclined to the rotation axis in the strongly-magnetic roAp stars, such that as the star rotates the pulsations are viewed from different orientations. Therefore, multi-night observations of candidate roAp stars are useful to ensure a confident (non)detection of high-frequency pulsation modes.

% - - - - - % - - - - - % - - - - - % - - - - - %
\begin{table*}
\caption{The three candidate roAp stars analysed in this work, with $B$- and $V$-mag values from \texttt{SIMBAD}, spectral type from \citet{Renson2009}, rotation periods from \citet{Bowman2018b} and magnetic field detection from \citet{Buysschaert2018b}.} 
\begin{center}
\begin{tabular}{l r r l r r}
\hline \hline
{Star name} & {$B$~mag} & {$V$~mag} & {Spectral Type} & {Rotation Period (d)} & {$B_{\rm polar}$ (kG)}\\\hline
HD~158596	&	9.17		&	8.94		&	B9 Si		&	$2.02208 \pm 0.00003$	&	1.8	\\
HD~166542	&	9.92		&	9.94		&	A0 Si		&	$3.6331 \pm 0.0003$	&	$-$	\\
HD~181810	&	10.65	&	10.66	&	A0 EuCrSr	&	$13.509 \pm 0.009$		&	$-$	\\
\hline \hline
\end{tabular}
\end{center}
\label{table: stars}
\end{table*}
% - - - - - % - - - - - % - - - - - % - - - - - %

% - - - - - % - - - - - % - - - - - % - - - - - %
\begin{table*}
\caption{Details of the observations of HD~158596, HD~166542 and HD~181810 obtained with the SHOC instrument mounted on the 1.0-m telescope at SAAO.}
\begin{center}
\begin{tabular}{l r r r r r r}
\hline \hline
\multicolumn{1}{c}{Star name} & \multicolumn{1}{c}{UTC} & \multicolumn{1}{c}{BJD -- 2450000} & \multicolumn{1}{c}{Length} & \multicolumn{1}{c}{Data points} & \multicolumn{1}{c}{Exp time} & \multicolumn{1}{c}{Observer} \\
\multicolumn{1}{c}{} & \multicolumn{1}{c}{} & \multicolumn{1}{c}{} & \multicolumn{1}{c}{(min)} & \multicolumn{1}{c}{} & \multicolumn{1}{c}{(sec)} & \multicolumn{1}{c}{} \\
\hline
HD~158596	&	2018-05-09	& 	8248.4139	& 	120.8	& 	1421		& 	5	& 	SBP		\\
			&	2018-05-15	& 	8254.4405	& 	120.7	& 	1430		& 	5	& 	SBP		\\
			& 	2018-06-13	&	8283.3179	&	146.9	&	862		&	10	&	DLH		\\
			& 	2018-06-16	&	8286.4243	&	220.4	&	1178		&	10	&	DLH		\\
			& 	2018-06-19	&	8289.5888	&	23.0		&	136		&	10	&	DLH		\\
\hline		
HD~166542	&	2018-05-09 	&	8248.5008	& 	120.8	& 	1428		& 	5	& 	SBP		\\
			&	2018-06-13	&	8283.5881	&	112.3	&	433		&	15	&	DLH		\\
\hline
HD~181810	&	2018-05-13 	& 	8252.4765	& 	310.1	& 	3624		& 	5	& 	SBP		\\
			&	2018-05-15	& 	8254.5273	& 	59.8		& 	718		& 	5	& 	SBP		\\
			&	2018-06-13	&	8283.4242	&	167.1	&	635		&	15	&	DLH		\\
\hline \hline
\end{tabular}
\end{center}
\label{table: observations}
\end{table*}
% - - - - - % - - - - - % - - - - - % - - - - - %

%	%	%	%	%	%	%	%	%	%

\subsection{HD~158596}
\label{subsection: HD158596}

\begin{figure}[t] 
\centerline{\includegraphics[width=0.99\columnwidth]{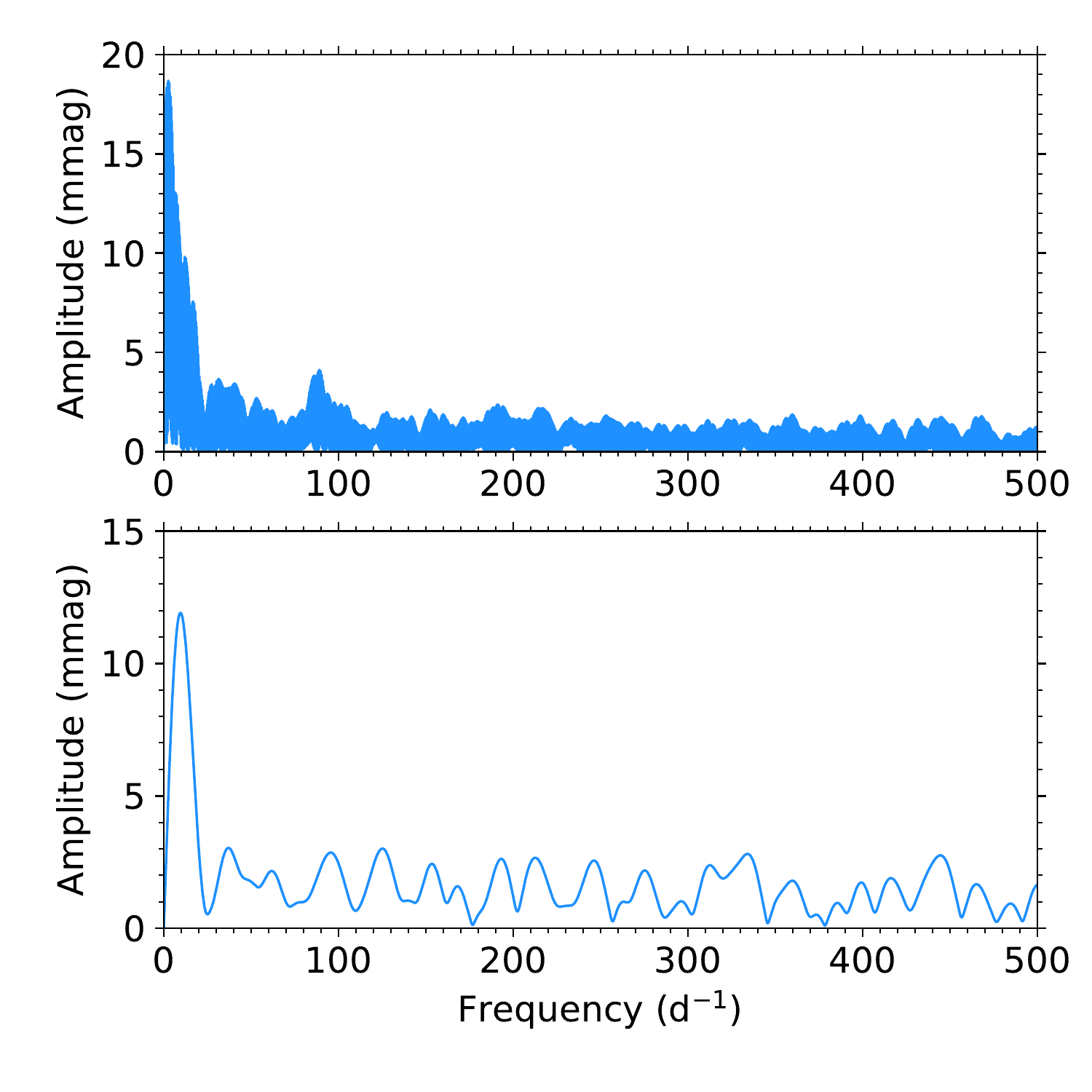}} 
\caption{The amplitude spectrum of the SAAO photometry of HD~158596 for all photometry (top panel) and the longest uninterrupted section (bottom panel).}
\label{figure: HD158596}
\end{figure}

Using K2 mission data from campaigns 9 and 11, \citet{Bowman2018b} determined a rotation period of $2.02208 \pm 0.00003$~d for HD~158596, and detected a significant isolated peak with a frequency of $17.0074 \pm 0.0004$~d$^{-1}$ in the amplitude spectrum. HD~158596 is also known to be a magnetic star with a large-scale magnetic field strength of 1.8~kG \citep{Buysschaert2018b}. The amplitude spectrum of the follow-up $B$ data obtained with SHOC/SAAO of HD~158596 is shown in Fig.~\ref{figure: HD158596}. Our data allow us to rule out high-amplitude high-frequency roAp pulsation modes in this star to a limit of $\sim$~1~mmag. It is known that there is a nearby star at a separation of 0.3~arcsec to HD~158596 \citep{ESA1997, Fabricius2002a}, but \citet{Buysschaert2018b} found no evidence of a secondary star in high-resolution and high-S/N spectroscopy. 

The follow-up high-cadence SHOC/SAAO data of HD~158596 span several days and cover a reasonable fraction of the rotation phase. From the lack of significant high-frequency variability above the noise level of $\sim$1~mmag, we conclude a null detection of high-frequency roAp pulsation modes in HD~158596. This implies that the frequency of $17.0074 \pm 0.0004$~d$^{-1}$ found by \citet{Bowman2018b} in K2 photometry is unlikely to be a Nyquist alias frequency of a high-frequency roAp pulsation mode frequency, and is possibly an intrinsic pulsation mode frequency or caused by a contaminating background object.
	
%	%	%	%	%	%	%	%	%	%

\subsection{HD~166542}
\label{subsection: HD166542}

\begin{figure}[t] 
\centerline{\includegraphics[width=0.99\columnwidth]{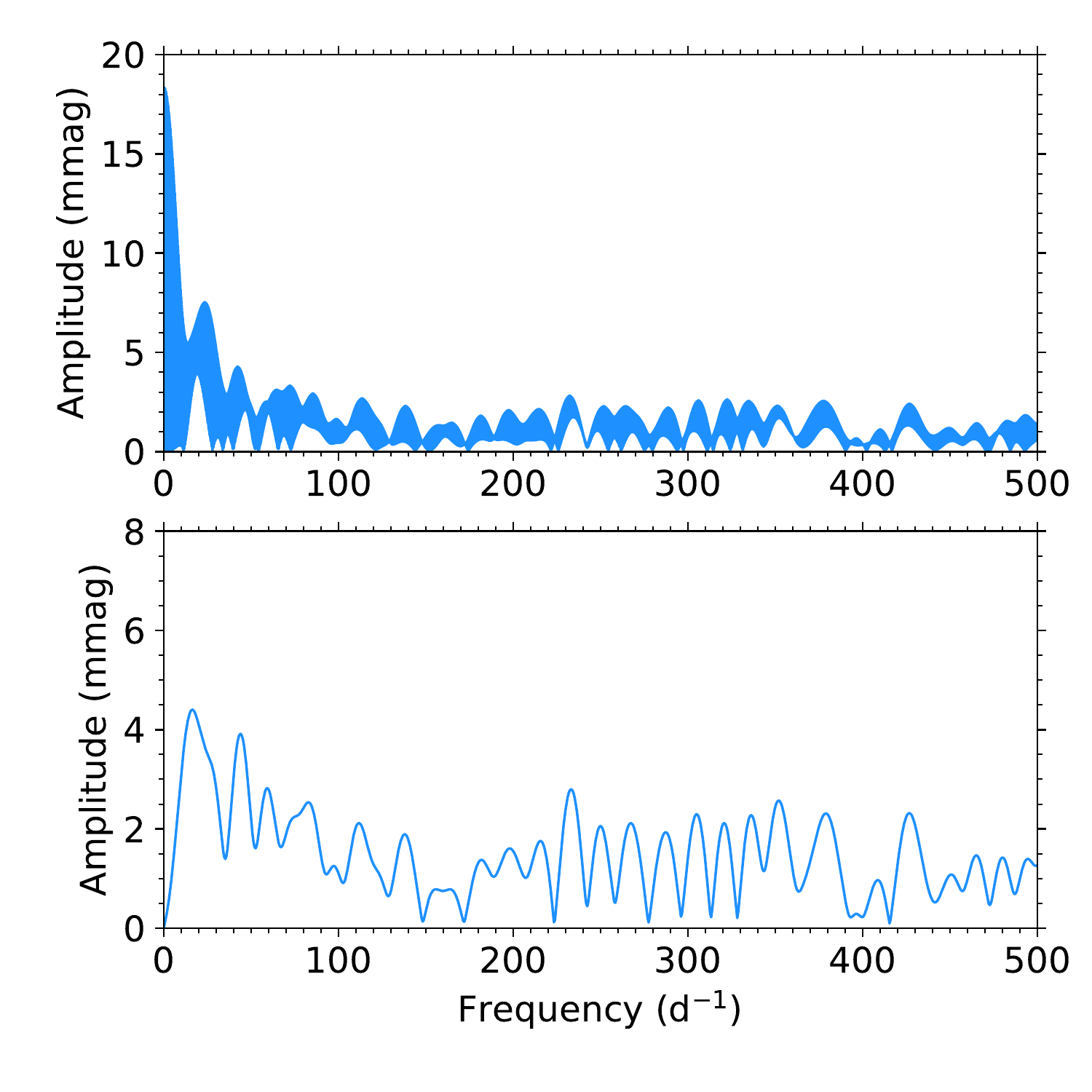}} 
\caption{The amplitude spectrum of the SAAO photometry of HD~166542 for all photometry (top panel) and the longest uninterrupted section (bottom panel).}
\label{figure: HD166542}
\end{figure}

For HD~166542, \citet{Bowman2018b} reported the first measurement of the rotation period of this star to be $3.6331 \pm 0.0003$~d using 70~d of K2 mission data from campaign 9, and found a significant isolated frequency peak at $16.429 \pm 0.004$~d$^{-1}$ in its amplitude spectrum. The amplitude spectrum of the follow-up $B$ data obtained with SHOC/SAAO of HD~166542 is shown in Fig.~\ref{figure: HD166542}. These data allow us to rule out high-frequency roAp pulsation modes in this star to a limit of $\sim1.4$~mmag. Thus, the frequency of $16.429 \pm 0.004$~d$^{-1}$ found by \citet{Bowman2018b} in K2 photometry is unlikely to be a Nyquist alias frequency, and may be caused by contamination from a background object or represent an intrinsic pulsation mode frequency.
	
%	%	%	%	%	%	%	%	%	%

\subsection{HD~181810}
\label{subsection: HD181810}

\begin{figure}[t] 
\centerline{\includegraphics[width=0.99\columnwidth]{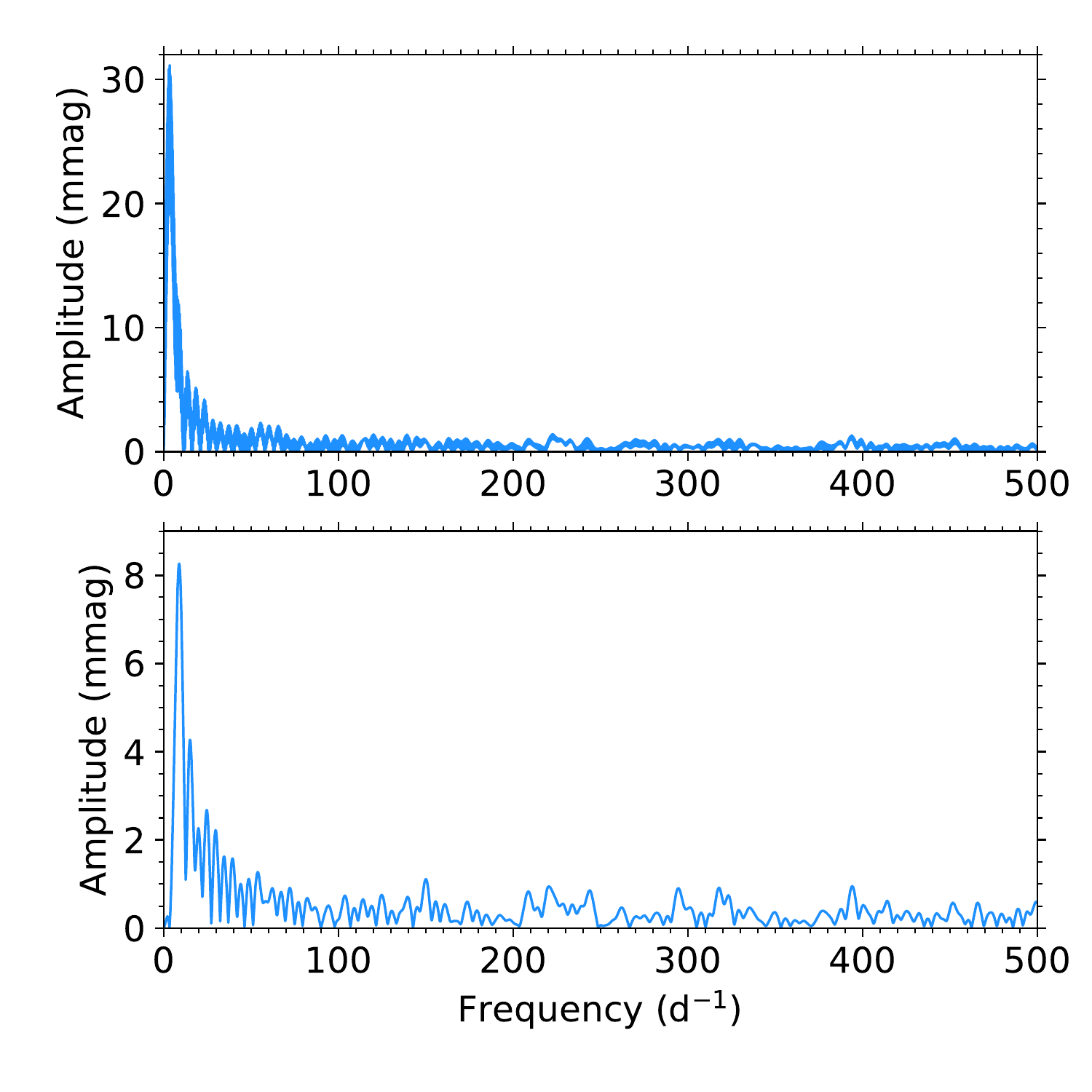}} 
\caption{The amplitude spectrum of the SAAO photometry of HD~181810 for all photometry (top panel) and the longest uninterrupted section (bottom panel).}
\label{figure: HD181810}
\end{figure}

HD~181810 was observed during campaign~7 of the K2 space mission, which \citet{Bowman2018b} used to measure the rotation period of this star to be $13.509 \pm 0.009$~d, but also detected a significant isolated frequency peak at $5.9331 \pm 0.0005$~d$^{-1}$ in its amplitude spectrum. The amplitude spectrum of the follow-up $B$ data obtained with SHOC/SAAO of HD~181810 is shown in Fig.~\ref{figure: HD181810}. These data allow us to rule out high-frequency roAp pulsation modes in this star to a limit of $\sim 0.6$~mmag. Since our observations cover a reasonable fraction of the rotation phase, we can exclude the detection of high-amplitude roAp pulsation modes in HD~181810. This null detection of suggests that the frequency of $5.9331 \pm 0.0005$~d$^{-1}$ found by \citet{Bowman2018b} is not a Nyquist alias frequency and might be caused by a contaminating background object or represent an intrinsic pulsation mode frequency.
	
% % % % % % % % % % % % % % % % % % % % % % % % % 

\section{Discussion}
\label{section: discussion}

In this paper, we have presented our interactive adaptive aperture photometry pipeline, {\sc TEA-Phot}, which is written in \texttt{PYTHON} and employs the functionality of the \texttt{SEP} package \citep{Barbary2016} to create customised light curves for the detection of pulsation modes in high-cadence ground-based photometry obtained with the SHOC instruments at SAAO. We have demonstrated that the use of adaptive elliptical apertures is preferable to circular apertures of fixed radii using the case study of the roAp star J1640, and the significant impact on the extraction of pulsation mode frequencies and amplitudes in variable stars. This is particularly true for observations obtained in non-photometric conditions. A significant advantage of the consequential lower average noise levels in the amplitude spectra of pulsating stars is that fewer observations are needed to detect high-amplitude pulsation modes.

The application of our {\sc TEA-Phot} pipeline to three candidate roAp stars, HD~158596, HD~166542 and HD~181810, using high-cadence ground-based $B$ data from the SHOC instrument mounted on 1.0-m telescope at SAAO yielded no detectable roAp pulsation modes in these stars at the noise level of order 1~mmag. Therefore, our results indicate that the low-frequency peaks discovered in the K2 mission data of these Ap stars are not Nyquist alias frequencies of roAp pulsation modes. On the other hand, if the frequencies detected by \citet{Bowman2018b} represent intrinsic pulsation modes then these ApBp stars pose an interesting challenge to pulsation theory since low overtone p~modes are not expected to be excited in the presence of a strong large-scale magnetic field \citep{Saio2005, Saio2014a}. 

Currently the TESS mission \citep{Ricker2015} is obtaining high-precision and high-cadence (i.e. 2-min) light curves of a subset of all known ApBp stars. TESS is observing almost the entire sky (i.e. $| b | > 6$~deg) and is providing light curves that range in length from 27~d up to 1~yr in its two continuous ecliptic viewing zones. Therefore, TESS is providing a rich and unique data set for finding new roAp stars and probing the largely unknown physics of how pulsation, rotation and strong magnetic fields interact within ApBp stars. We refer the reader to \citet{David-Uraz2019b} and \citet{Cunha2019a} for recent TESS results of rotation and pulsation detected in ApBp stars in the southern ecliptic hemisphere using TESS mission data. For stars located in the ecliptic plane, however, the K2 mission data will remain an important legacy sample of variable stars for the foreseeable future. With the increased chance of finding more candidate roAp stars at an increasing rate, it is more important than ever before to have an efficient pipeline such as {\sc TEA-Phot} available to analyse follow-up observations of these stars, particularly for stars that are not observed by the nominal TESS mission at a high cadence.

%%%%%%%%%%%%%%%%%%%%%%%%%%%%%%%%%%%%%%%%%%%%%%%%%%

\begin{acknowledgements}
The authors are grateful to Stephen B. Potter for obtaining some of the observations used in this work, and DMB would like to thank Jels Boulangier for useful discussions. The research leading to these results has received funding from the European Research Council (ERC) under the European Union's Horizon 2020 research and innovation programme (grant agreement No.~670519: MAMSIE). DLH acknowledges the Science and Technology Facilities Council (STFC) via grant ST/M000877/1 and the National Research Foundation (NRF) of South Africa. This research has made use of the \texttt{SIMBAD} database, operated at CDS, Strasbourg, France; the SAO/NASA Astrophysics Data System; and the VizieR catalogue access tool, CDS, Strasbourg, France. This research has made use of \texttt{ASTROPY}, a community-developed core \texttt{PYTHON} package for Astronomy \citep{Astropy2013}; and \texttt{MATPLOTLIB}, a \texttt{PYTHON} library for publication quality graphics \citep{Hunter2007}; and \texttt{SEP}, a source extraction and aperture photometry package \citep{Barbary2016} based on the Source Extractor package \citep{Bertin1996}.
\end{acknowledgements}

%%%%%%%%%%%%%%%%%%%%%%%%%%%%%%%%%%%%%%%%%%%%%%%%%%

\bibliographystyle{aa}
\bibliography{/Users/Dominic/Documents/RESEARCH/Bibliography/master_bib}

%%%%%%%%%%%%%%%%%%%%%%%%%%%%%%%%%%%%%%%%%%%%%%%%%%

\end{document}